\documentclass[fleqn,10pt]{wlscirep}
\usepackage[utf8]{inputenc}
\usepackage[T1]{fontenc}
\usepackage{hyperref}
\usepackage{authblk}
\usepackage{xr}
\usepackage{array,enumitem,makecell}
\usepackage{pgfplots}
\usepackage{pgfplotstable,siunitx,booktabs,colortbl}
\pgfplotsset{compat=1.17}

\title{AI-driven materials design: a mini-review}

\author[1,2,3,$\dagger$,*]{Mouyang Cheng}
\author[1,4,$\dagger$]{Chu-Liang Fu}
\author[1,5,$\dagger$]{Ryotaro Okabe}
\author[1,6,$\dagger$]{Abhijatmedhi Chotrattanapituk}
\author[1,4]{Artittaya Boonkird}
\author[7]{Nguyen Tuan Hung}
\author[1,2,4,**]{Mingda Li}
\affil[1]{Quantum Measurement Group, MIT, Cambridge, MA 02139, USA}
\affil[2]{Center for Computational Science $\&$ Engineering, MIT, Cambridge, MA 02139, USA}
\affil[3]{Department of Materials Science and Engineering, MIT, Cambridge, MA 02139, USA}
\affil[4]{Department of Nuclear Science and Engineering, MIT, Cambridge, MA 02139, USA}
\affil[5]{Department of Chemistry, MIT, Cambridge, MA 02139, USA}
\affil[6]{Department of Electrical Engineering and Computer Science, MIT, Cambridge, MA 02139, USA}
\affil[7]{Frontier Research Institute for Interdisciplinary Sciences, Tohoku University, Sendai, 980-8578, Japan}
\affil[$\dagger$]{These authors contributed equally.}
\affil[*]{e-mail: vipandyc@mit.edu}
\affil[**]{e-mail: mingda@mit.edu}

\begin{abstract}
Materials design is an important component of modern science and technology, yet traditional approaches rely heavily on trial-and-error and can be inefficient. 
Computational techniques, enhanced by modern artificial intelligence (AI), have greatly accelerated the design of new materials.
Among these approaches, inverse design has shown great promise in designing materials that meet specific property requirements.
In this mini-review, we summarize key computational advancements for materials design over the past few decades. 
We follow the evolution of relevant materials design techniques, from high-throughput forward machine learning (ML) methods and evolutionary algorithms, to advanced AI strategies like reinforcement learning (RL) and deep generative models.
We highlight the paradigm shift from conventional screening approaches to inverse generation driven by deep generative models. Finally, we discuss current challenges and future perspectives of materials inverse design. This review may serve as a brief guide to the approaches, progress, and outlook of designing future functional materials with technological relevance.

\end{abstract}
\begin{document}

\flushbottom
\maketitle

\thispagestyle{empty}

\section{Introduction}
Materials design plays an indispensible role in modern technology, from steel in the industrial revolution to semiconductors powering information age.
Traditional approaches of materials discovery rely on iterative and resource-intensive searches, and are often constrained by human intuition and experimental limitations. 
These methods, while successful in the past century, struggle to meet the growing demand for materials with highly specialized properties for modern technology, such as energy materials, catalysis materials, and quantum materials.
Around the turn of the millennium, computational techniques, such as hierarchically structured materials design\cite{olson1997computational} and integrated computational materials engineering (ICME)\cite{olson2000designing}, transformed materials discovery by using computational models to automatically identify promising material candidates.
Since then, high-throughput computational methodologies have flourished for materials design. Leveraging quantum mechanical and molecular dynamics simulations, these methods enable systematic exploration of a broader chemical space. Moreover, with the advent of AI, they can be further accelerated by surrogate models that can directly predict materials' properties.

However, most high-throughput approaches  follow a ``forward screening'' paradigm, where materials are first generated, then filtered based on a target property. 
Such forward-screening approach faces huge challenges in materials design, where the chemical and structural design space is astronomically large\cite{oganov2019structure}.
Identifying thermodynamically stable materials with superior properties remains difficult in such a huge design space.
The stringent conditions for stable materials design result in high failure rates in na\"ive traversal approaches, making forward screening highly inefficient.
By contrast, ``inverse design'' reverts this paradigm, starting from targeting properties and designing desirable materials backward\cite{zunger2018inverse,han2024ai}. This approach holds immense promise for discovering materials with superior target properties for specific applications.
Early attempts such as genetic algorithms (GAs) and Monte Carlo tree search (MCTS) laid the groundwork for inverse design development\cite{oganov2006crystal,woodley2008crystal}. However, they often fall short in performance and efficiency when scaled up to the vast and complex design space, due to the pre-defined exploration paths with limited flexibility.

\begin{figure}[!htbp]
    \centering
    \includegraphics[width=0.9\linewidth]{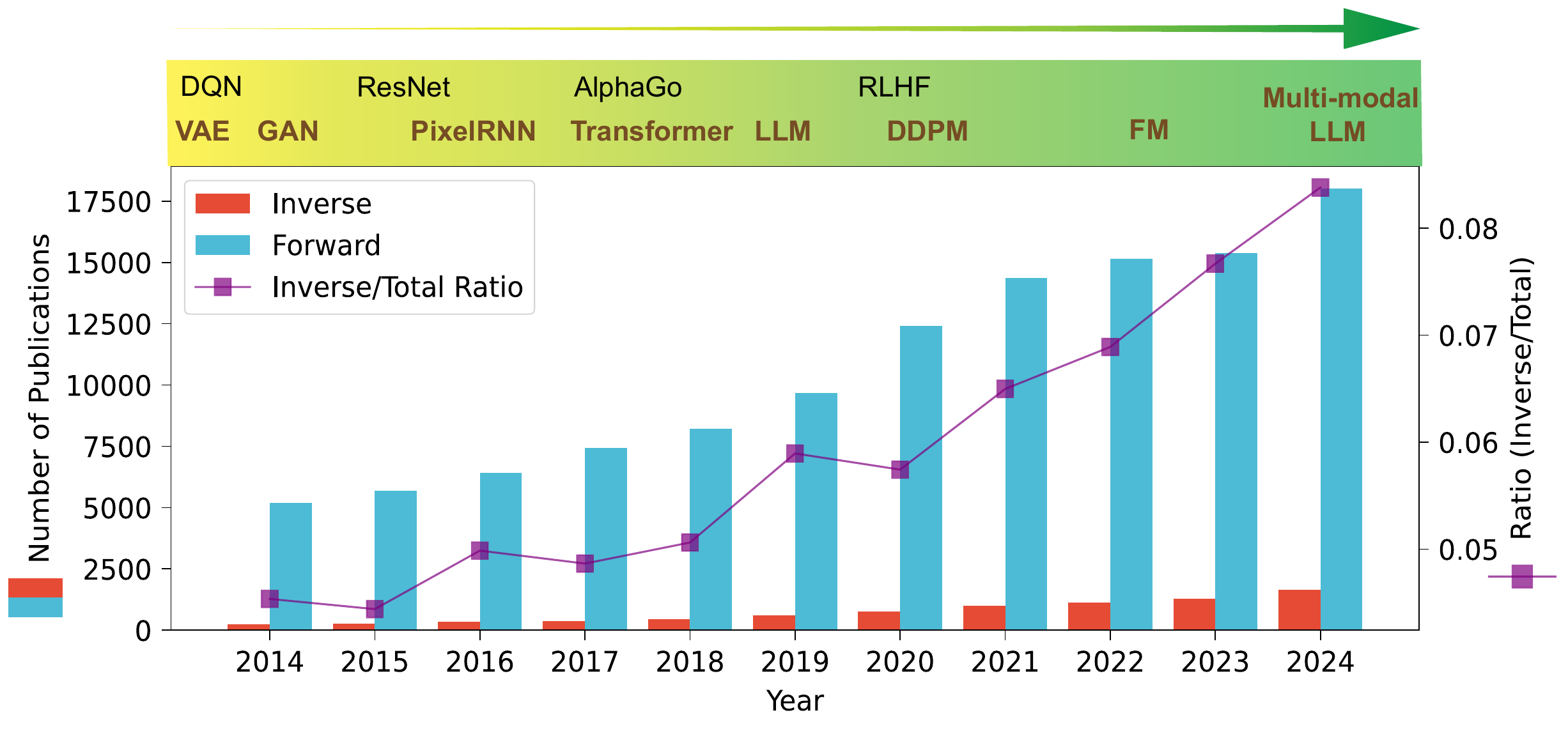}
    \caption{\textbf{Evolutionary trend of materials design paradigm for forward vs. inverse design since 2014, by number of publications.} The publications presented in this figure are retrieved from the Web of Science database with keywords related to automatic materials design and discovery. 
    Publications with inverse design are filtered based on additional keywords with various AI-based inverse-design techniques. %Only publications involving deep generative models are classified as inverse design.
    Milestone AI techniques and frameworks are labeled on the top panel along the time line, with deep generative models marked bold in brown.}
    \label{fig1}
\end{figure}

In the past decade, AI powered by deep neural networks has emerged as a transformative tool in materials design. 
Deep reinforvement learning (RL) methods such as deep Q-learning  \cite{mnih2013playing} and RL with human feedback (RLHF) \cite{christiano2017deep} introduce real-time feedback on the computational output, which could optimize the efficiency of inverse design.
Moreover, deep generative models such as variational autoencoders (VAEs)\cite{kingma2013auto}, generative adversarial networks (GANs)\cite{goodfellow2014generative}, deep autoregressive models\cite{van2016pixel}, and diffusion models\cite{ho2020denoising} have drastically reshaped the landscape of inverse design.
The power of these deep generative models lies in their robust ability to map the intricate relationships between materials' structures and properties, and directly perform materials generation conditioned on target properties. 
For instance, latent space-based methods enable the navigation of high-dimensional design spaces, while diffusion models utilize stochastic processes to facilitate the generation of novel material candidates with desired properties. 
Thus, these conditional generative models are well suited for inverse design, as they are capable of efficiently learning and sampling from the vast and nonlinear chemical spaces in materials.
To provide a more quantitative view on the research paradigm of materials design, Fig.\,\ref{fig1} illustrates the evolutionary trend in the field and highlights key advancements in AI since 2014. 
One can clearly observe from Fig.\,\ref{fig1} a growing trend of inverse design, now accounting for approximately 8\% of the materials design literature.
This reveals a paradigm shift from traditional ``forward design'' approaches to ``inverse design'' methodologies, driven by the powerful AI techniques. 

This mini-review provides a brief overview on the progress in AI-driven materials design. We structure this review as follows. Section 2 covers forward-screening design methods. Section 3 discusses early classical algorithms, including GAs, particle swarm optimization (PSO), and MCTS. Section 4 introduces more advanced adaptive and interactive methods such as Bayesian optimization (BO) and RL. Given the central role of experiments in materials science, we also introduce autonomous laboratories. Section 5 explores cutting-edge deep generative models, including VAEs, GANs, and more recently diffusion models for stable material generation, and large language models (LLMs). Finally, Section 6 discusses current challenges and future directions in integrating inverse design into scientific research and technological innovation, outlining a potential roadmap for accelerated materials discovery. A few notable computational methods related to materials design are summarized in Table.\,\ref{tab1}. 

% the table is large, so move to table1.tex
\definecolor{rulecolor}{RGB}{0,71,171}
\definecolor{tableheadcolor}{gray}{0.92}
% Following is taken from Werner: http://tex.stackexchange.com/a/33761/3061
% and modified for my needs
%
% Command \topline consists of a (slightly modified)
% \toprule followed by a \heavyrule rule of colour tableheadcolor
% (hence, 2 separate rules)
\definecolor{rulecolor}{RGB}{0,71,171}
\definecolor{tableheadcolor}{gray}{0.92}

% Define custom rules and styles
\newcommand{\topline}{ %
        \arrayrulecolor{rulecolor}\specialrule{0.1em}{\abovetopsep}{0pt}%
        \arrayrulecolor{tableheadcolor}\specialrule{\belowrulesep}{0pt}{0pt}%
        \arrayrulecolor{rulecolor}}
\newcommand{\midtopline}{ %
        \arrayrulecolor{tableheadcolor}\specialrule{\aboverulesep}{0pt}{0pt}%
        \arrayrulecolor{rulecolor}\specialrule{\lightrulewidth}{0pt}{0pt}%
        \arrayrulecolor{white}\specialrule{\belowrulesep}{0pt}{0pt}%
        \arrayrulecolor{rulecolor}}
\newcommand{\bottomline}{ %
       \addlinespace[-0.8ex]%      
      \arrayrulecolor{white}\specialrule{\aboverulesep}{0pt}{0pt}%
        \arrayrulecolor{rulecolor} %
        \specialrule{\heavyrulewidth}{0pt}{\belowbottomsep}}%

% Custom section header command
\newcommand{\topmidheader}[2]{%
    \addlinespace[0.6ex]% Adds space before the header row
    \multicolumn{#1}{c}{\textsc{#2}}\\%
    \addlinespace[1ex]% Adds space after the header row
}
\newcommand{\midheader}[2]{%
    \addlinespace[-0.6ex]%
        \midrule\topmidheader{#1}{#2}}
\setlist[itemize]{noitemsep, topsep=0pt, partopsep=0pt, leftmargin=1em}
\renewcommand{\arraystretch}{0.8}

\begin{table}[!ht]
    \centering
    \begin{tabular}[t]{p{3.5cm} c p{3.5cm} p{6cm}} % Added 'p{6cm}' for wrapping text
        \topline
        \rowcolor{tableheadcolor}
        \textbf{Method} & \textbf{Original year} & \textbf{Essence} & \textbf{Pros \& Cons} \\
        \midtopline
        \topmidheader{4}{Evolutionary algorithms}
        Genetic algorithm (GA)  & 1973  & Evolutionary search based on natural selection principles &
        \vspace{-0.6em}\begin{itemize}
            \item Intuitive; robust to noisy and multi-modal problems
            \item May converge prematurely to suboptimal solutions
        \end{itemize}
        \\
        Particle swarm optimization (PSO)          & 1995 & Swarm intelligence inspired by birds' flocking behavior &\vspace{-0.6em}\begin{itemize}
            \item Efficient for continuous optimization problems
            \item Heavily depends on parameter tuning
        \end{itemize}\\
        Monte Carlo tree search (MCTS)       & 2006  & Probabilistic decision-making using random sampling in trees &\vspace{-0.6em}\begin{itemize}
            \item Balances exploration and exploitation
            \item Computationally intensive for large problem spaces
        \end{itemize}\\
        \midheader{4}{Adaptive and interactive learning methods}
        Bayesian optimization (BO)         & 1978   & Sequential inference for global optimization of black-box functions&\vspace{-0.6em}\begin{itemize}
            \item Adaptive and data-efficient
            \item Computationally intensive; heavily rely on choices on prior
        \end{itemize}\\
        Deep reinforcement learning (DQN, etc.)  & 2013 & Neural networks approximating reward functions &\vspace{-0.6em}\begin{itemize}
            \item Learns complex policies from raw data
            \item Inefficient training; hard hyperparameter tuning
        \end{itemize}\\
        \midheader{4}{Deep generative models}
        Variational autoencoder (VAE) & 2013      & Probabilistic latent space learning via variational inference&\vspace{-0.6em}\begin{itemize}
            \item Effective for generative modeling
            \item Variational assumption limits expressiveness
        \end{itemize}\\
        Generative adversarial network (GAN)  & 2014      & Adversarial learning between generator and discriminator&\vspace{-0.6em}\begin{itemize}
            \item Generates highly realistic data
            \item Training is unstable; prone to mode collapse.
        \end{itemize}\\
        Large language model (LLM)  & 2017      & Transformer-based pretraining for language understanding&\vspace{-0.6em}\begin{itemize}
            \item Exceptional in natural language tasks and versatile domains
            \item Requires enormous computational resources; biased output
        \end{itemize}\\
        Diffusion model  & 2020      & Progressive noise removal to generate data&\vspace{-0.6em}\begin{itemize}
            \item High-quality and stable outputs
            \item Slow generation process; requires careful tuning.
        \end{itemize}\\
        \bottomline
    \end{tabular}
    \caption{An incomplete list of classical algorithms and AI-based computational methods for materials inverse design.}
    \label{tab1}
\end{table}

\section{Forward Screening}\label{fwd}
\begin{figure}[!htbp]
    \centering
    \includegraphics[width=0.95\linewidth]{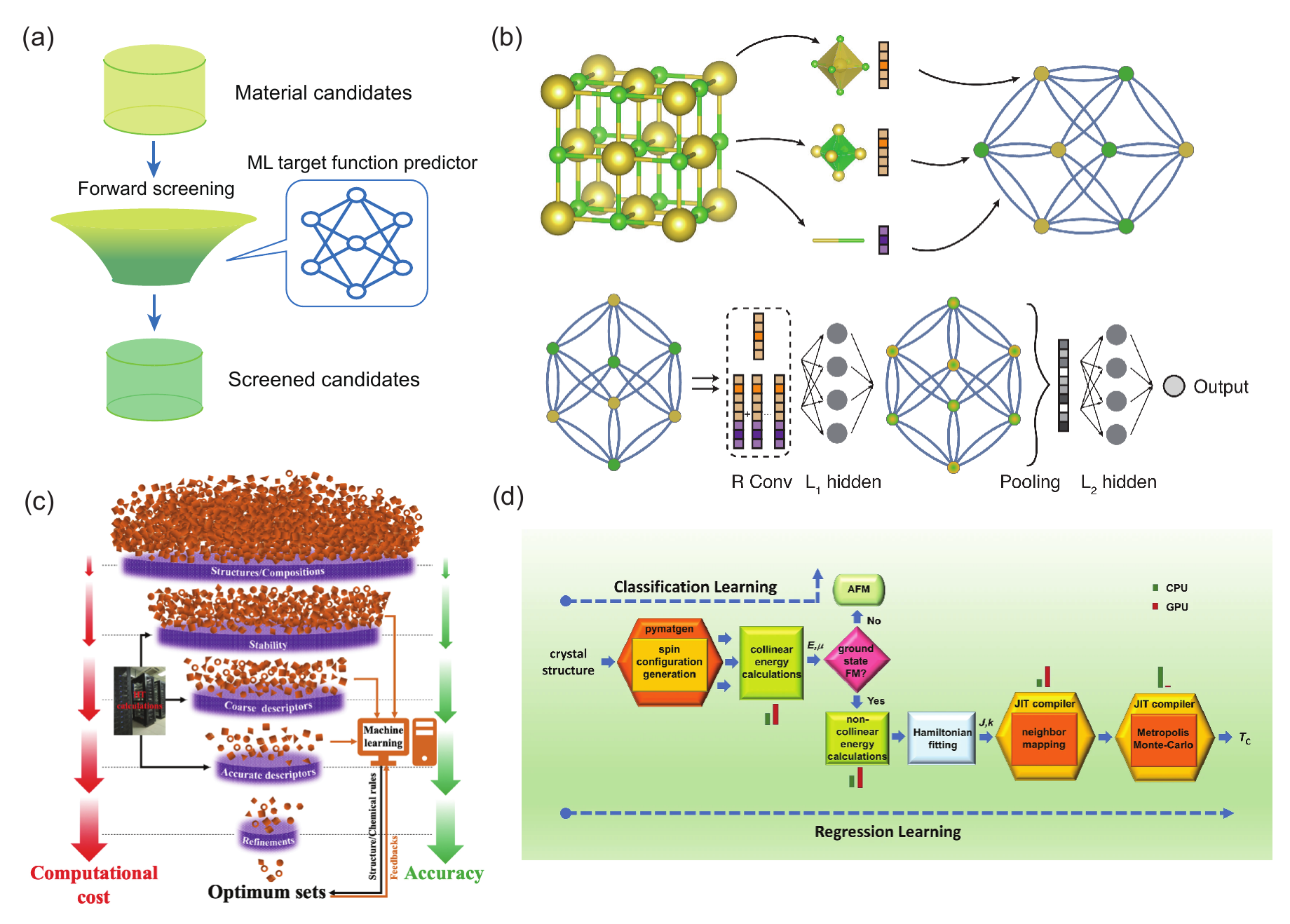}
    \caption{\textbf{Material discovery based on forward screening.} \textbf{(a)} Schematics of forward screening, which filters material candidates by high-throughput calculation of target properties. Such calculations can be accelerated by ML surrogate models. \textbf{(b)} Crystal graph convolutional neural networks (CGCNN) as a prominent framework of GNN for materials property prediction and screening\cite{xie2018crystal}. \textbf{(c)} A five-stage high-throughput computational screening process (purple disks) for optoelectronic semiconductors\cite{luo2021high}. Candidates are filtered based on stability and descriptor-based refinements. Machine learning speeds up predictions while balancing computational cost (red arrows) and accuracy (green arrows). \textbf{(d)} The screening workflow for identifying 2D ferromagnetic (FM) materials\cite{kabiraj2020high}. Spin configurations undergo energy calculations to classify materials as FM or antiferromagnetic (AFM). FM candidates are further refined through Hamiltonian fitting and Monte Carlo simulations to estimate the Curie temperature. Reproduced with permission\cite{xie2018crystal,luo2021high,kabiraj2020high}.}
    \label{fig2}
\end{figure}

We begin with forward screening, a natural and widely used methodology in computational materials discovery. This method systematically evaluates a set of predefined materials to identify those that meet the target property criteria.
Fig.\,\ref{fig2} shows the workflow and examples of materials design with the forward screening approach. As shown in Fig.\,\ref{fig2}(a), the forward screening typically starts from collecting candidate materials, often sourced from open-source databases\cite{jain2013commentary}. 
With the database properly set up, the forward screening process involves setting property thresholds based on domain-specific requirements. These thresholds act as filters to systematically eliminate candidates that do not meet the desired specifications. 
Automated frameworks such as Atomate\cite{mathew2017atomate} and AFLOW\cite{curtarolo2012aflow} greatly facilitate such forward screening processes by streamlining data preparation,  density functional theory (DFT) calculations, and post-analysis.
Conventionally, first-principles methods like DFT are often employed to compute various materials properties. 
However, the significant computational cost of these methods makes it impractical to directly screen a very large dataset. To this end, ML surrogate models provide a powerful alternative, accelerating the evaluation process by predicting properties with much lower computational costs. In many workflows, computationally intensive screening methods are reserved at the final stage to ensure the reliability while maintaining the overall efficiency.

Another critical aspect of forward screening is the representation of materials, which significantly influences the accuracy of property predictions. Numerous structural representations are proposed for materials, and summarized in a few reviews\cite{musil2021physics,reiser2022graph}.
Among them, graph neural networks (GNNs) excel for their ability to capture the geometric features of material structures. As a prominent example, GNNs can represent the atomistic systems by representing the atoms as the graph nodes and the interatomic relations as the edges\cite{xie2018crystal} (Fig.\,\ref{fig2}(b)), though it is worth noting that graph nodes have a broader meaning beyond atomic representations\cite{Okabe2024VGNN}. 

Forward screening has been extensively applied across materials with various target functions, including stability, electronic, thermal and magnetic properties. For example, it has been used to screen materials based on thermodynamic stability, in order to identify compounds that can exist under specific conditions, such as energy above the convex hull and phonon dispersion computed by machine learning interatomic potentials\cite{batatia2023foundation}.
It has also been employed to discover new optoelectronic semiconductors, including light absorbers, transparent conductors, light-emitting diode (LED) materials, and photovoltaic materials, by screening for electronic properties\cite{luo2021high} (Fig.\,\ref{fig2}(c)). 
In thermal applications, high-throughput screening has been performed to identify half-heusler semiconductors with extremely low thermal conductivity for heat management systems\cite{carrete2014finding}. In addition, forward screening has facilitated magnetic materials design with applications such as permanent magnets, spintronics, and magnetic refrigeration\cite{zhang2021high}. Besides traditional bulk materials, forward screening methods have been applied to two-dimensional (2D) materials. For example, it has been used to develop an automated and high-throughput workflow to screen 2D ferromagnetic materials. Combining DFT and Monte Carlo simulations, 2D ferromagnetic materials with a Curie temperature above 400 K were identified \cite{kabiraj2020high} (Fig.\,\ref{fig2}(d)). Given the vast literature, the examples here only represent a small fraction of the extensive efforts in forward screening approaches.

Despite its achievements, forward screening faces a few fundamental challenges. One major challenge is the lack of exploration: All screening workflows operate as one-way process, applying criteria on existing databases, without the capability to extrapolate beyond known data distributions. This poses a challenge for forward screening to discover novel materials with properties beyond existing trends. 
Another fundamental limitation is the severe class imbalance -- only a small fraction of candidates exhibit desirable properties, leading to a very low success rate of screening. As a result, the majority of the computational resources can be spent on evaluating materials that ultimately fail to meet the target criteria. These limitations make forward screening far from ideal for modern computational materials design, highlighting the urgent need for inverse design algorithms operating under an inherently different paradigm.

\section{Evolutionary Algorithms (EAs)}\label{evo}
The early approaches of materials inverse design are based on evolutionary algorithms (EAs), a class of global optimization methods inspired by natural evolution principles. Materials candidates are often represented as a set of parameters defining structures and properties, making it challenging to find the optimal combinations simultaneously. EAs evaluate the parameter sets through a fitness function as the metrics quantifying the performances with respect to the specific design objectives, including catalytic performance, hardness, synthesizability, and magnetization\cite{le2016discovery, allahyari2020coevolutionary}. The algorithm explores the design space by promoting beneficial traits and discarding less effective ones, thereby mimicking the process of biological evolution. While the computational costs increase as we explore the broad range of candidate spaces, these adaptive search strategies are well-suited for the complex, high-dimensional landscapes encountered in multi-scale optimization in the materials design process.

\begin{figure}[!htbp]
    \centering
    \includegraphics[width=0.9\linewidth]{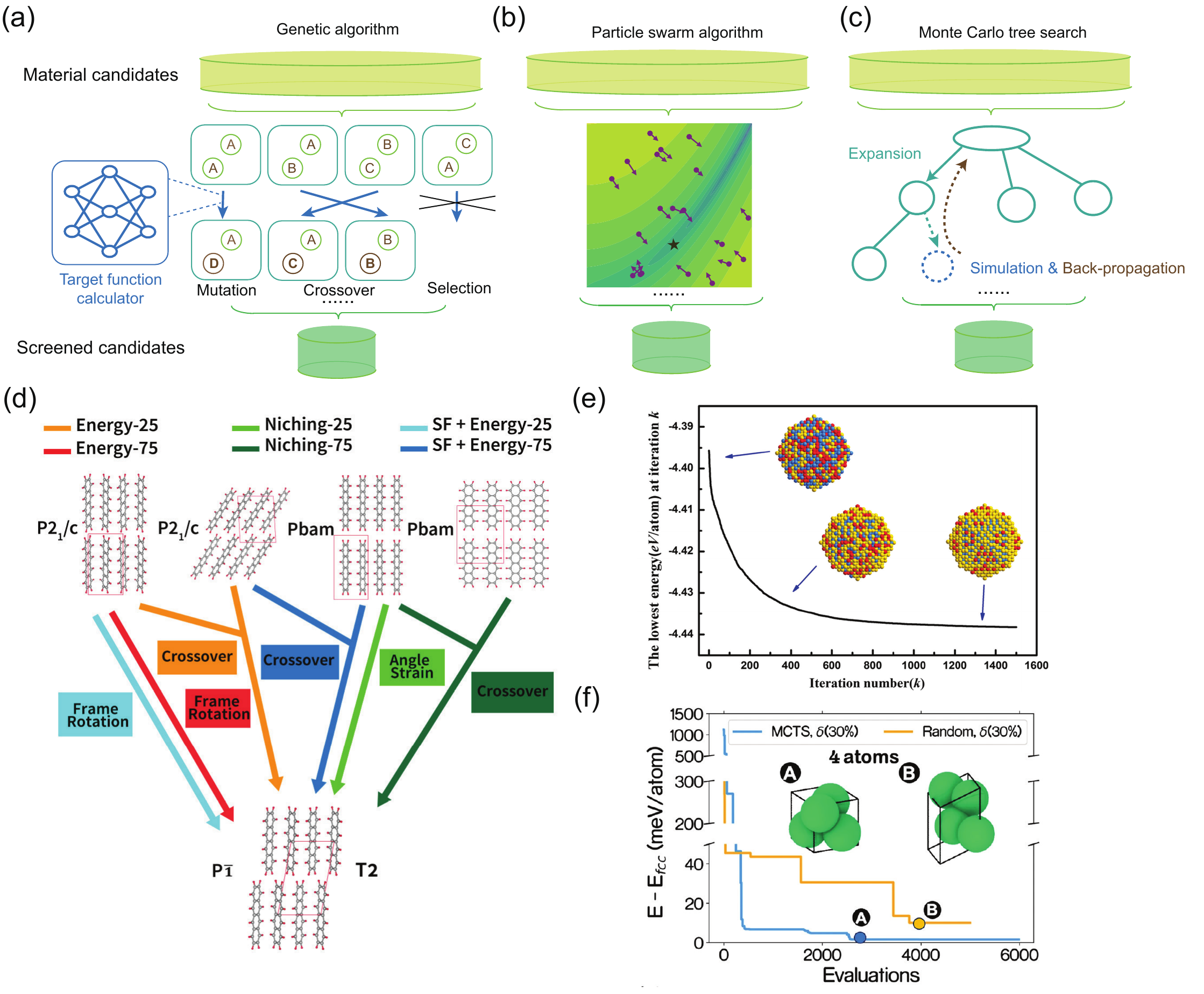}
    \caption{\textbf{Evolutionary algorithms (EAs) for materials inverse design.} The top diagrams show the schematics of \textbf{(a)} genetic algorithm (GA), where candidate materials undergo mutation (A to D), crossover (B \& C exchanged), and selection driven by a target function calculator; \textbf{(b)} Particle swarm optimization (PSO), where candidate solutions evolve by following individual (purple arrows) and the global optimum position (star) in a search space; and \textbf{(c)} Monte Carlo tree search (MCTS), where a decision tree is expanded (green solid arrow) iteratively using simulations (blue dashed circle) and back-propagation (brown dashed arrow) to identify optimal candidates. \textbf{(d)} Evolutionary routes that explore the candidate structures of tetracene in GA\cite{tom2023inverse}. Different evolutionary strategies, including crossover, frame rotation, and angle strain, guide structural transformations between space groups, and various selection criteria (Energy-based, Niching, and SF+Energy) influence the optimization pathway. \textbf{(e)} The evolutionary process of the stable structure for Pt–Pd–Au nanoparticles with 3285 atoms in PSO\cite{fan2015structural}, with corresponding nanoparticle structures shown at a three different evolutionary stages. \textbf{(f)} Comparison of energy convergence between MCTS (blue curve) and random sampling (orange curve) of a four-atom silver cluster (green spheres). MCTS reaches faster convergence and identifies an optimal structure with lower-energy (label ``A'') compared to random sampling (label ``B'')\cite{banik2023continuous}. Reproduced with permission\cite{tom2023inverse,fan2015structural,banik2023continuous}.}
    \label{fig3}
\end{figure}

EAs can be categorized into methods such as GAs, PSO, and MCTS, each with its unique approach to solving non-linear, high-dimensional, and multimodal optimization problems. GAs (Fig.\,\ref{fig3}(a)) mimic the natural selection process by encoding candidate materials and employing crossover, mutation, and selection operations to evolve the population over successful generations. GAs are particularly effective in discrete optimization problems, where materials are represented as a set of structural parameters encoded in genetic representations. In contrast, PSO (Fig.\,\ref{fig3}(b)) draws inspiration from the collective behavior of social orgasms, treating each set of candidates as particles following the optimization trajectory in the design space based on their behavior and those of their neighbors. Its ability to efficiently traverse high-dimensional landscapes makes it well-suited for tuning material properties as continuous variables. On the other hand, MCTS (Fig.\,\ref{fig3}(c)) incorporates the elements of randomness and systematic search by exploring the decision trees to find the optimal solutions. The tree-based search strategy prioritizes promising solution pathways, making it particularly useful in combinatorial problems. The following sections describe the roles and applications of GAs, PSO, and MCTS in materials discovery.

Among the various EAs, GAs are among the most widely applied in materials design. GAs maintain a population of candidate solutions and refine them through iterative selection, mutation, and crossover processes. These steps mimic natural selection, where the fitter candidates are more like to be chosen for the next generation. GAs have been particularly effective in crystal structure prediction, where they identify stable atomic configurations by minimizing the total energy of candidate structures. For example, GA has been combined with DFT to predict an Fe$_2$P phase as the first post-pyrite phase of SiO$_2$ at low temperatures\cite{wu2011identification}.
Property-based GAs have been implemented to inverse-design polymorphic crystal structures of tetracene, using a characteristic-based fitness function to identify structures with higher singlet fission (SF) performance and stability\cite{tom2023inverse} (Fig.\,\ref{fig3}(d)). 
GAs themselves can also be accelerated using surrogate ML models. For instance, a machine learning-accelerated genetic algorithm (MLaGA) was developed, integrating an on-the-fly Gaussian process with GA. This approach reduced the number of DFT calculations by up to 50-fold compared to traditional GA approaches and was used to identify the optimal chemical ordering within binary alloy nanoparticles\cite{jennings2019genetic}.

PSO is a prominent EA inspired by the social behavior of birds or fish. In PSO, candidate solutions, represented as particles, iteratively adjust their positions in the search space based on both their own experiences and those of their neighbors. This cooperative approach allows PSO to explore complex design space efficiently and converge to global optima. PSO has been widely applied in materials design, such as thermal or electronic properties, optimizing complex structures where traditional methods might falter. For example, PSO has been used as a crystal structure predictor, which is implemented in the CALYPSO package\cite{wang2012calypso}. The CALYPSO code efficiently optimizes the free energy surface and generates crystal structures based solely only on chemical composition information\cite{wang2010crystal}. By incorporating symmetry constraints and eliminating structural overlaps, PSO improves computational efficiency and has successfully predicted various new high-pressure phases of lithium and silica. PSO has also been used for designing materials with superior mechanical properties, where it was used to construct a hardness v.s. energy map to predict energetically favorable superhard structures. Benchmarked on carbon, B-N, and B-C-N systems, PSO reproduced both experimentally and theoretically known structures\cite{zhang2013first}. An improved discrete PSO algorithm was introduced to investigate the structural stability and surface segregation behavior of tetrahexahedral Pt–Pd–Au trimetallic nanoparticles\cite{fan2015structural} (Fig.\,\ref{fig3}(e)). By integrating swap operators and exchange probabilities, this method effectively optimized the nanoparticles' configuration with Pt occupying interior regions and Pd and Au segregating to the surface.

MCTS is a probabilistic algorithm that employs random sampling to explore decision trees. In materials design, MCTS has shown promise in identifying optimal configurations for complex material systems. By systematically sampling from possible coordination states and prioritizing promising branches, MCTS facilitates the efficient exploration of combinatorial search spaces. 
MCTS has been applied to identify stable structures, such as the segregation of silver impurities in copper, by searching only 1\% of all the possible configurations\cite{kiyohara2018searching}. Another framework called Continuous Action Space Tree search for INverse desiGn (CASTING) has been proposed for materials inverse design, employing MCTS for continuous action space to navigate a complex energy landscape\cite{banik2023continuous} (Fig.\,\ref{fig3}(f)). The framework demonstrates scalability and accuracy across diverse materials systems, including metals, covalent systems, and complex oxides, with applications in discovering superhard carbon phases and optimizing multi-objective properties. MCTS further contributes to solving complex material design problems like alloy systems\cite{m2017mdts}. The structural optimization and composition constraints of silicon-germanium (Si-Ge) alloys have been achieved to achieve minimum and maximum thermal conductance.

Despite their successes, EAs still face significant challenges. One major limitation is their computational cost, as the iterative evaluation of candidate solutions often requires the computation of many configurations, which can be computationally prohibitive for large datasets. Premature convergence is another common issue, where the algorithm is trapped in local optima, missing other optimized solutions elsewhere in the search space. Also, generalization is a common challenge for EA-based inverse design, as many EAs are applied to the systems with predefined chemical composition. Moreover, the performance of EAs is highly sensitive to parameter choices, including population size, mutation rate, and selection pressure. The absence of standardized optimization techniques makes tuning these parameters a non-trivial task, often requiring extensive trial and error. This challenge further complicates the application of EAs in materials inverse design.

\section{Adaptive and Interactive Approaches}\label{rl}
In contrast to EAs, which heavily rely on stochastic exploration, another category of methods is
\textit{adaptive} and \textit{interactive}, which dynamically 
 updates search strategies based on prior knowledge and feedback. This section explores how such methods, such as BO, RL, and autonomous lab, can be used for inverse design of materials. 
 
\begin{figure}[!htbp]
    \centering
    \includegraphics[width=1\linewidth]{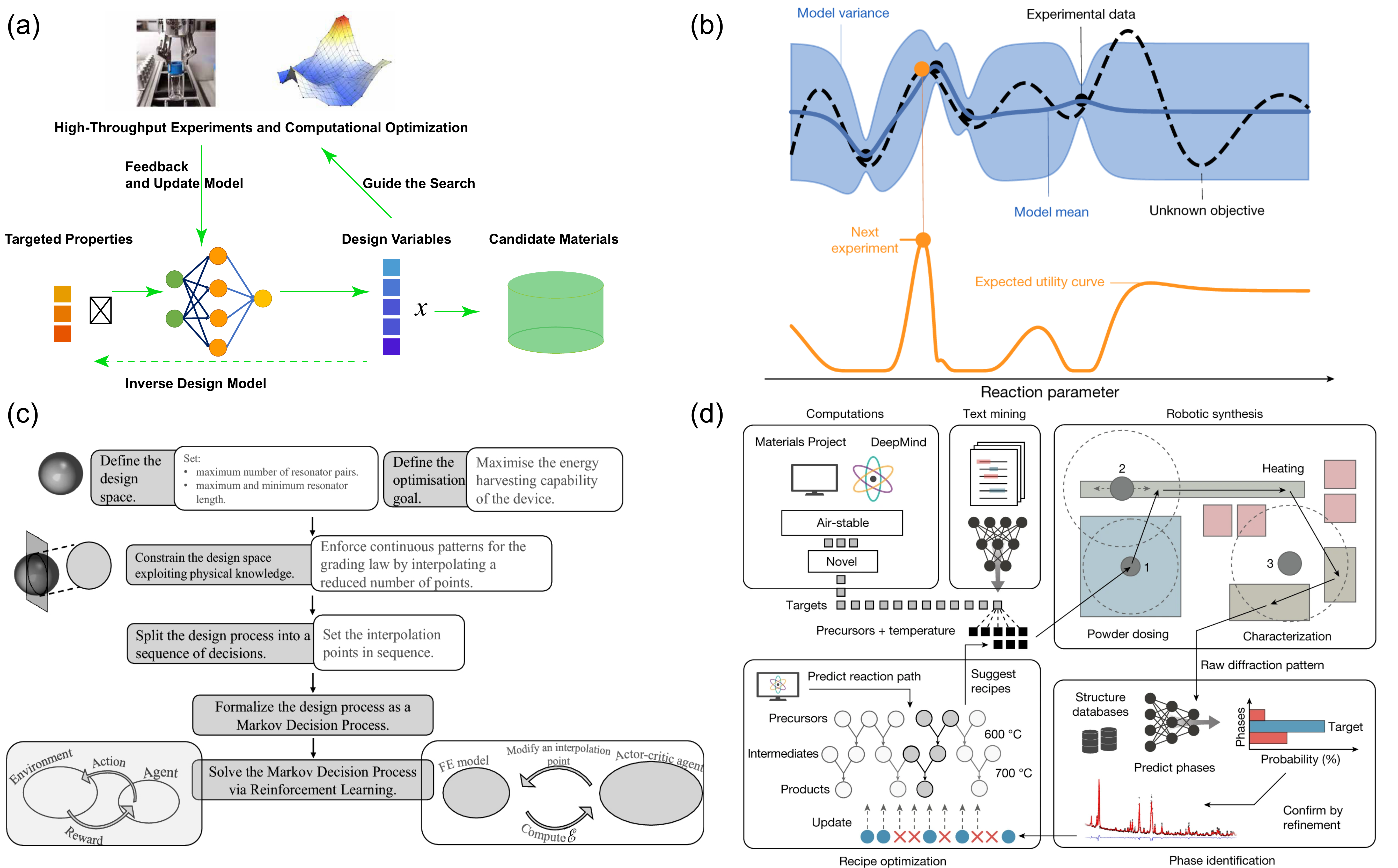}
    \caption{
    \textbf{Adaptive and interactive approaches for inverse design. (a)} A schematic to illustrate the adaptive approach, where the inverse design model is associated with the guidance and feedback/update loops to the experiments and computation. \textbf{(b)} Graphical illustration of the BO used for chemical reaction \cite{shields2021bayesian}. A Gaussian process surrogate model is fitted to data from an unknown objective (black dashed line), with the posterior mean (blue curve) shown with a shaded $2\sigma$ region. The expected utility curve (orange curve) guides the next experiment, which updates the current knowledge of the surrogate model. \textbf{(c)} Methodology illustration to design optimization of graded metamaterials for energy harvesting\cite{rosafalco2023reinforcement}. The design process is framed as a Markov Decision Process, where physical constraints and interpolation points guide sequential decisions, and is iteratively optimized through reinforcement learning. \textbf{(d)} The illustration of autonomous materials discovery with the A-Lab \cite{szymanski2023autonomous}. A-Lab integrates computation (top left), text mining (top middle), robotic synthesis (right), recipe optimization (bottom left), and phase identification (bottom right) to optimize reaction pathways. Machine learning suggests synthesis recipes, experimental robots execute synthesis and characterization, and diffraction analysis confirms phase identification. Reproduced with permission\cite{shields2021bayesian,rosafalco2023reinforcement,szymanski2023autonomous}. }
    \label{fig4}
\end{figure}

Adaptive inverse design involves refining a design model by iteratively updating it based on the results from forward simulations or experimental data. In each iteration, the model generates candidate materials that progressively approach the desired property targets, guided by feedback from prior evaluations. The interaction feedback loop between the design model and materials property predictions or evaluation is typically realized through forward calculations or experiments, informed by computational simulations, empirical datasets, or automated experiments. A schematic illustration of adaptive inverse design is presented in Fig.\,\ref{fig4}(a).
A key advantage of this inverse design paradigm is its compatibility with high-throughput computational screening and experimental platforms, both of which enable real-time data generation and adaptive model refinement. This dynamic feedback loop accelerates the exploration of material spaces by continuously improving the predictive accuracy and efficiency of the design model. When integrated with experimental validation, the synergy between data-driven computational guidance and physical realization has significantly advanced the discovery of novel materials, contributing to the rapid progress observed in recent years.

To achieve optimal performance in adaptive inverse design, specialized AI frameworks must be employed to accelerate the discovery process while maintaining scientific rigor. We introduce three foundational adaptive and interactive approaches that achieve inverse design: BO, a probabilistic optimization framework that balances exploration and exploitation, making it well-suited for adaptively refining material candidates based on limited data; RL, a ML paradigm increasingly enhanced by deep learning, capable of navigating complex design spaces through an agent that interacts with environment and receives reward; Laboratory automation, advanced experimental automation tools that enable autonomous high-throughput experimentation and data collection, ensuring efficient feedback between physical validation and computational modeling. The integration of these AI-driven methodologies holds great promise for transforming materials discovery by bridging computational predictions with experimental validation in a dynamic, iterative design framework.

BO is a powerful framework for global optimization in scenarios where data is scarce or expensive to acquire \cite{shahriari2015taking}. It constructs a probabilistic surrogate model, often a Gaussian process, to approximate the objective function. The surrogate model estimates the uncertainty in the objective function and balances exploration (sampling new areas of the design space) with exploitation (focusing on areas likely to yield optimal results). Key techniques such as acquisition functions, including expected improvement (EI), probability of improvement (PI), and upper confidence bound (UCB), are used to guide the next set of experiments by selecting points that maximize the improvement potential. BO has been successfully applied in materials discovery for tuning synthesis or processing conditions with minimal experimental runs.

A promising application of BO lies in reaction optimization \cite{shields2021bayesian}, shown in Fig.\,\ref{fig4}(b). Here the 1D Gaussian process surrogate model for reaction parameters is constructed and fitted to data collected from an unknown objective and the corresponding expected utility curve (surface for high dimensional case). The model is plotted as the posterior mean with 2$\sigma$ variance presented as the shaded region. The expected utility curve is maximized to select the next experiment to iteratively update the model until completely optimized. The employment of this reaction BO demonstrates superiority over human decision-making in both efficiency and consistency. Additionally, BO was further applied to synthetically valuable optimization problems, including the Mitsunobu and deoxyfluorination reactions, showcasing its versatility in chemical synthesis. BO is increasingly employed as an adaptive surrogate model for automated experimental design \cite{lei2021bayesian}. Recent developments include multi-scale batch BO for efficient navigation of microstructure spaces under optimized processing conditions \cite{honarmandi2022accelerated} and a variant capable of handling missing data when target materials fail to form due to suboptimal growth parameters \cite{wakabayashi2022bayesian}.

RL combines concepts from ML and optimal control to solve sequential decision-making problems \cite{mnih2013playing,christiano2017deep}. RL models contain an agent that interacts with an environment, where the agent takes actions based on observations and receives rewards that reflect performance. The goal is to maximize cumulative rewards over time by learning an optimal policy. 

In materials design, RL can be applied to explore complex, high-dimensional design spaces by learning policies that maximize a reward function, such as achieving a specific property threshold. Advanced algorithms like Q-learning, deep Q-networks (DQN), policy gradient methods, and actor-critic frameworks, have enabled RL to guide autonomous materials synthesis while adapting to experimental feedback dynamically. 
Recent RL applications in inverse materials design include the compositional design of multicomponent alloys \cite{xian2024compositional}, high thermal conductivity amorphous polymers \cite{ma2022exploring}, and the design of mechanical metamaterials with nonlinear deformation responses \cite{brown2023deep}. Although RL often requires large datasets, its data demands can be mitigated by incorporating physical constraints and tailored algorithm design. For instance, RL has been successfully applied to optimize the design of graded metamaterials \cite{rosafalco2023reinforcement} for energy harvesting under realistic conditions, such as magnetic loading and random excitations (Fig.\,\ref{fig4}(c)). In this work,  materials design space is formulated into a decision sequence by setting the interpolation points in sequence, and the decision process can be treated as a Markov decision process, which is solved through RL. In gray background regions, the steps are applied to formalize and solve a general design problem with RL. In white background regions, the corresponding operations are applied to this concrete design of graded metamaterials. The results emphasize RL's capability to manage design tasks involving high-dimensional, stochastic optimization problems while revealing unexpected, physically meaningful design patterns. It is worth emphasizing that this work leverages physical insights to reduce the computational complexity. This physics-informed approach enables the RL agent to operate within a reduced design space and may inspire further exploration to impose other physics constraints RL for inverse design. 

Apart from computational breakthroughs, recent advances in laboratory automation have also significantly contributed to the acceleration of materials design and discovery. Automated systems, such as robotic arms, can perform high-throughput synthesis and characterization, executing hundreds to thousands of experiments with minimal human intervention. Techniques such as automated synthesis, X-ray and neutron spectroscopy, and electron microscopy, generate vast datasets essential for AI models. Laboratory automation platforms can integrate with AI frameworks to perform closed-loop optimization, where experimental conditions are adjusted automatically based on real-time, on-the-fly feedback from computational models. By coupling laboratory automation with real-time data analysis and feedback loops, it has enabled rapid validation of AI-driven predictions and the discovery of novel materials. While experimental assessments provide essential data that feed into the computational models, the computational models, in turn, refine their predictions based on this data, updating parameters to offer improved guidance for subsequent experiments. 
A recent notable advancement in autonomous materials discovery is the A-Lab, \cite{szymanski2023autonomous} as shown in Fig.\,\ref{fig4}(d) for its materials-discovery pipeline with the integration of computations, text mining, robotic synthesis, recipe optimization, and phase identification. The A-Lab represents an autonomous laboratory designed to accelerate the synthesis of novel inorganic materials by seamlessly integrating ML algorithms and advanced robotics. Its key innovation lies in the fully automated workflow, including precursor handling, robotic synthesis, and X-ray diffraction, all controlled by a closed-loop system where experimental outcomes are continuously analyzed and optimized using active learning algorithms. Other notable efforts combining experimental techniques with inverse design include but are not limited to the quantum dot synthesis \cite{epps2020artificial} and AlphaFlow, an RL-based platform optimized for the synthetic pathways in the shell-growth of core-shell semiconductor nanoparticles \cite{volk2023alphaflow}. These works further highlight the synergy between automation, machine learning, and experimental methodologies in advancing materials design.

Despite its great potential, the adoption of adaptive inverse materials design remains limited for several reasons. Firstly, forward calculations often lack the speed required for real-time model updates, creating a bottleneck in the iterative feedback loop. Secondly, the current materials science community tends to favor the construction of extensive static databases that aim to cover entire design spaces, rather than engaging in dynamic model refinement during the discovery process. Finally, implementing such an adaptive framework requires additional infrastructure and methodological efforts, including the development of workflows that seamlessly connect forward exploration calculations with inverse design models. These challenges underscore the necessity of reviewing the current state of the field and encouraging further research into more effective implementations of interactive inverse materials design.
Nevertheless, by employing these three approaches: BO, RL, and laboratory automation (which can be guided by BO or RL), we believe that AI-driven adaptive design has the potential to make inverse materials design more efficient than ever before.

\section{Deep Generative Models}
Deep generative models represent the state-of-the-art approach for materials inverse design.
Widely known for the recent popularization of realistic image, text, audio, and video generation, deep generative models' essence is to approximate the probability distribution of the training dataset. Due to the probabilistic nature, these models sample data points according to the learned distribution and are capable of creating novel outputs, while maintaining coherence and realism. Within the scope of materials inverse design, deep generative models can be trained on known materials that are either experimentally synthesized or computationally optimized, to approximate the realistic material distribution in the vast chemical design space. From there, two levels of integration can be achieved for the inverse design pipeline.  
The first level generates stable materials without conditional bias, and the second level is a conditional generation, from which the model is guided to directly generate materials with desired properties. Although all deep generative models perform the similar task of approximating data distribution, many techniques have been developed since the early 2010s, including VAE, GAN, diffusion model and LLM.

\subsection{Variational Autoencoders (VAEs)}
A VAE is a generative model that learns to encode the distribution of the training dataset into a predetermined latent space distribution, from which new data can be sampled and decoded to generate\cite{kingma2013auto}. As a derivative of autoencoder, a VAE consists of two sub-modules: encoder and decoder. In a regular autoencoder, the encoder learns to \textit{deterministically} map the input data into a low dimensional latent space, while the decoder learns to do the reverse by constructing the original data from the latent representation. These two sub-modules are trained jointly, allowing the reconstruction of the input data from the latent representation. Hence, the autoencoder usually runs for dimensional reduction or feature extraction of the training data. However, the standard autoencoder does not guarantee the semantic continuity of the latent space. For instance, the neighborhoods of training data points in the latent space do not guarantee similar decoded outputs as the original data, including falling out of the data distribution or even being unrealistic. Furthermore, the learned latent distribution can be of irregular shape, making the sampling process difficult to perform. 

\begin{figure}[!htbp]
    \centering
    \includegraphics[width=0.95\linewidth]{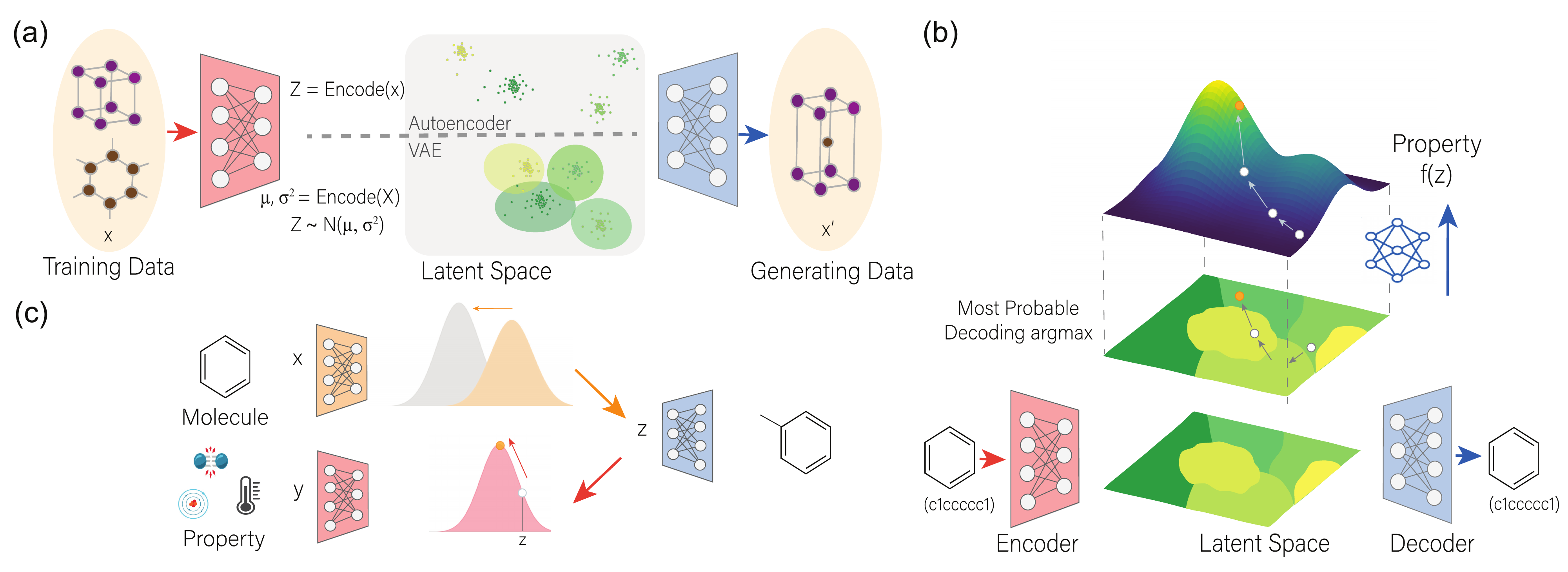}
    \caption{\textbf{Variational autoencoder (VAE) approaches for inverse design.} \textbf{(a)} The general idea behind the autoencoder (AE) and VAE approach: In an AE, the encoder maps inputs to deterministic points in latent space $Z$, while in a VAE, the latent space $Z$ is constrained by a probabilistic prior (e.g., Gaussian distribution). Compared to AE, the structured latent space of VAE enhances semantic continuity and can be used for data generation. \textbf{(b)} A common approach for inverse design with VAE via latent space search and optimization. The latent representation of each input structure can be trained with the corresponding properties by another small machine learning model to create a property map, in which the search and optimization algorithm can provide candidate latent vectors, from which the decoder can reconstruct them to obtain structures with the desired properties\cite{gomez2018automatic}. \textbf{(c)} Jointly trained VAE between structure-property learning\cite{fallani2024inverse}. The model aligns latent spaces ($z$) of molecular structures ($x$) and properties ($y$), enabling direct inverse mapping from properties (bottom left) to target structures (right) for efficient materials design. Reproduced with permission\cite{gomez2018automatic,fallani2024inverse}.}
    \label{fig5}
\end{figure}

VAE solves these problems with probabilistic encoding and latent distribution enforcement (Fig.\,\ref{fig5}(a)). In a VAE, the encoder maps each input to a distribution (typically Gaussian) instead of a point in latent space. This means that the decoder needs to learn to reconstruct the same input from multiple neighboring points in the latent space, which drives the latent space to be continuous. VAE also regulates the learned latent distribution to be as close as possible to a predefined distribution, which allows the generation of new, realistic data by directly sampling from latent space.

While VAE was originally designed for sampling new realistic data from the probability distribution of the training dataset, its structured latent space makes it particularly well-suited for materials inverse design. By mapping complex, high-dimensional material data to a lower-dimensional latent space, VAE enables efficient exploration and optimization of material properties within the latent space. The decoder, in turn, serves as a generator to reconstruct material structures or compositions based on desired features or target performance from the latent space. This framework not only accelerates the discovery of novel materials but also facilitates the incorporation of domain-specific constraints with the aid of latent space.

Recently, several explorations have been made along this track. One of the early attempts comes from Gómez-Bombarell \textit{et al.} to develop VAE to generate novel molecules for drug and materials design \cite{gomez2018automatic}. The input molecule information is represented using SMILES (Simplified Molecular Input Line Entry System) strings, a discretized textual format that encodes molecular structures with ASCII characters. With VAE, this discrete representation is transformed into a continuous latent space, which allows further operations like decoding random vectors, perturbing known chemical structures, or interpolating between molecules, to facilitate tasks like molecular optimization and novel molecule generation. While the interest in discovering new molecules usually requires maximizing some desirable property, this generative model also predicts property values from the latent representation by connecting to another neural network for property prediction, shown in Fig.\,\ref{fig5}(b). In the figure, the additional property mapping in latent space allows molecule generation with the targeted properties through ML. The same strategy with VAE can also be used for polymer solar cells search \cite{jorgensen2018machine} with SMILES mapping, alloy microstructure inverse design \cite{pei2021machine}, and nanoporous crystalline reticular materials \cite{yao2021inverse}, etc. Some modifications of the typical VAE architecture have also been developed to enhance its capability. One recent work encodes molecule structure and property jointly to reach a common latent representation for both properties and structures \cite{fallani2024inverse}. As a result, combining the property encoder with the decoder of VAE can achieve direct inverse mapping from the property to the candidate structure, as shown in Fig.\,\ref{fig5}(c). While the encoder in orange embeds the molecule structure information, the encoder in red embeds the property information to connect the property-structure relation and enable direct inverse mapping.

Although VAE can generate new data from the same distribution as the training dataset, and enable properties exploration and optimization in latent space, it faces challenges in capturing the complex distributions in the latent space. VAE usually lacks fine-tuning and requires to balance two competing objectives, the reconstruction of input data from latent representation, and enforcing the regularization to construct a structured latent space distribution to ensure smoothness and meaningful interpolation during the training process. This regularization usually constrains the model’s ability to retain fine details and results in a low-quality reconstruction. Therefore, VAE has been gradually replaced by or supplemented by other generative models, such as diffusion models.

\subsection{Generative adversarial networks (GANs)}
A GAN consists of two sub-modules: a generator and a discriminator. The generator learns to create realistic data, while the discriminator learns to distinguish generated data from the real data. With this adversarial mechanism, GANs implicitly model data distributions without assuming prior distributions, offering flexibility in generating novel and realistic structures\cite{goodfellow2014generative}. In materials design, the generator learns to create realistic crystal structures by capturing underlying structural and compositional rules, while the discriminator ensures the adherence to these rules. GANs have demonstrated success in materials generation during the past few years. For example, a GAN was developed for Mg-Mn-O ternary systems, achieving notable success in discovering new compositions and predicting their stability through high-throughput virtual screening\cite{kim2020generative}. Another deep convolutional GAN links crystal images with structural and formation energy constraints, enabling formation-energy-guided structure generation\cite{long2021constrained}.  GAN can also be used in VAE latent space sampling, as demonstrated in perovskite material generation. Instead of directly using a latent vector, letting the generator propose a better, this approach allows the generator to refine nearby latent vectors under lattice constraints, which addresses VAE's low-quality reconstruction problem that can lead to materials with lower symmetries\cite{Chenebuah2024}.

However, one major disadvantage of GAN is the difficulty of training. GAN is known to suffer from the instability issues such as mode collapse, where the generator fails to capture fully the data distribution. This could partially explain why GANs have received less attention in materials inverse design compared to other generative models.

\subsection{Diffusion models}

\begin{figure}[!htbp]
    \centering
    \includegraphics[width=0.9\linewidth]{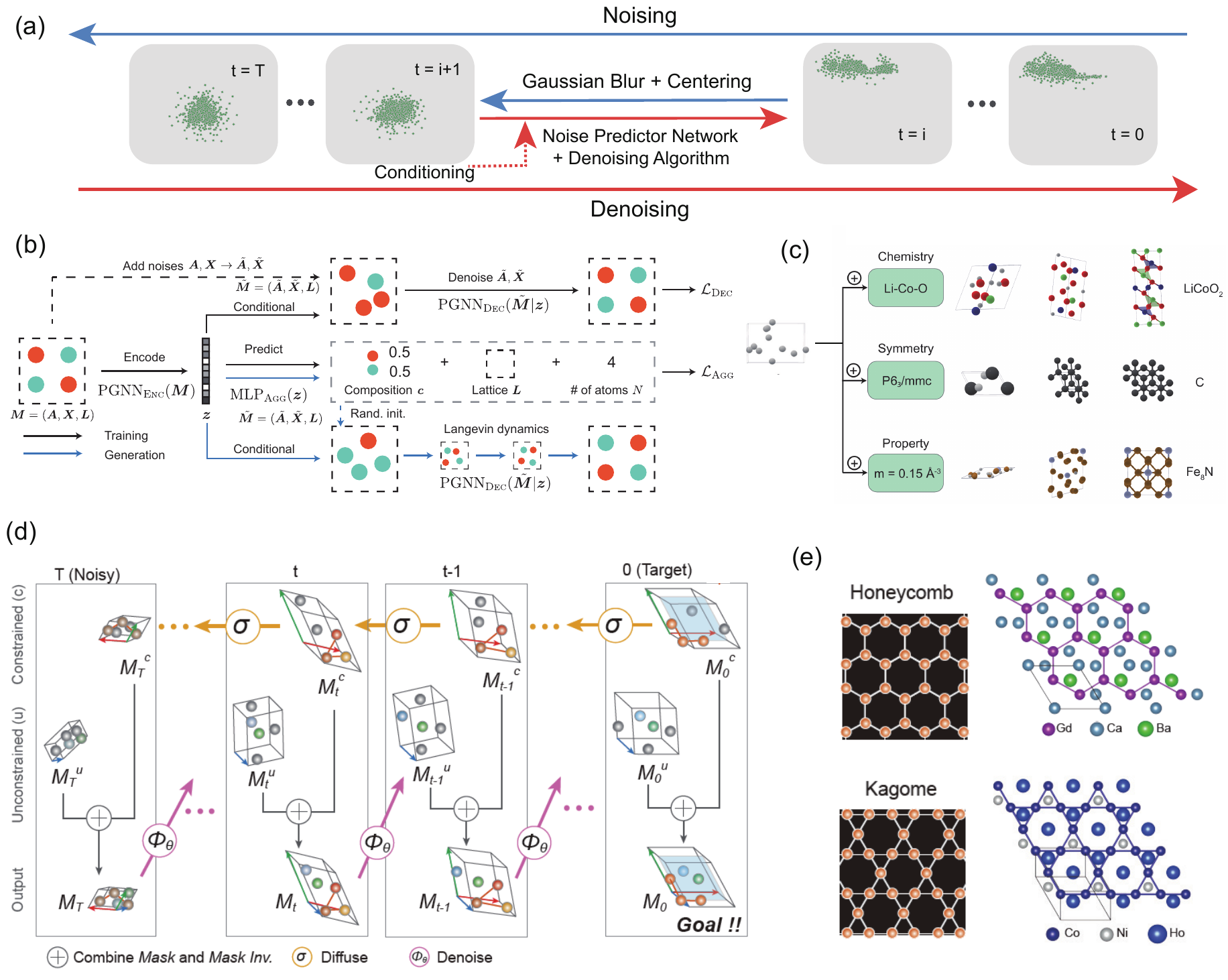}
    \caption{\textbf{Diffusion model approaches for inverse design.} \textbf{(a)} The general workflow of a diffusion model, which consists of noising and denoising steps. The noise is gradually introduced as Gaussian blur from $t=0$, ensuring that the distribution approaches a standard Gaussian at the final timestep $t=T$. Starting from $t=T$, the model is trained to reverse each noising step (optionally with conditional bias) to steer the denoising into the original distribution at $t=0$. \textbf{(b)} Crystal diffusion variational autoencoder (CDVAE) architecture combining diffusion model and VAE for material generation\cite{xie2021crystal}. The VAE block generates the crystal lattice $\mathbf{L}$, initial chemical composition and number of atoms, then the diffusion blocks jointly updates atomic types $\mathbf{A}$ and coordinates $\mathbf{X}$ to refine a crystal structure. The latent space representation can also be used for property-guided conditional generation. \textbf{(c)} MatterGen's conditional generation of materials with fine-tuning of the diffusion model\cite{zeni2025generative}. The foundational model can be fine-tuned to generate materials with various constraints, including chemical elements, geometric symmetry and material properties. \textbf{(d)} Structural constraint integration in generative model (SCIGEN) for imposing motif geometries as constraints to the crystal structure generation process\cite{okabe2024structural}.
    For each denoising step $t$, an unconstrained structure $\boldsymbol{M}_{t}^u$ is combined with constrained structure $\boldsymbol{M}_{t}^c$ to get an integrated structure $\boldsymbol{M}_{t}$. $\boldsymbol{M}_{t}$ is passed to the denoising model $\Phi_{\theta}$ and denoised to become the unconstrained structure $\boldsymbol{M}_{t-1}^u$. By repeating this process, the final crystal structure $\boldsymbol{M}_{0}$ is obtained under the geometrical pattern constraints $\boldsymbol{M}_{0}^c$.
    \textbf{(e)} The examples of patterns (Honeycomb and Kagome) and generated crystal structures applied in SCIGEN. Reproduced with permission \cite{xie2021crystal,zeni2025generative,okabe2024structural}.}
    \label{fig6}
\end{figure}

Diffusion models represent an emerging class of state-of-the-art generative models that learn data distributions through a progressive noising and denoising process. Inspired by non-equilibrium thermodynamics, these models learn to reverse the gradual addition of noise to a dataset, reconstructing the original data while preserving the underlying structure and key properties. Common architectures like denoising diffusion probabilistic models (DDPM) and score-based diffusion models (SBDM) have shown performance in generating complex data, including images, texts, and high-dimensional tensors. Training a diffusion model involves minimizing a loss function derived from the reconstruction loss and noise prediction loss, allowing the model to approximate the true data distribution effectively while learning to refine noisy samples. The processes are visualized in Fig.\,\ref{fig6}(a).

Diffusion models have been applied to materials science problems. For instance, a video-based diffusion model was introduced for the inverse design of nonlinear mechanical metamaterials, enabling the prediction of stress-strain responses over nonlinear deformation paths. This approach bypasses traditional mappings from properties to designs by training on full-field deformation data. Instead, it predicts the expected deformation path and internal stress distributions, achieving remarkable agreement with finite element simulations\cite{bastek2023inverse}. Diffusion models have also been employed for alloy design and optimization. Denoising diffusion models work to generate novel alloy compositions and optimize their properties across multiple generative tasks. The probabilistic nature of these models not only accounts for solution non-uniqueness but also facilitates uncertainty quantification, enabling robust exploration of complex composition-property relationships\cite{fernandez2024denoising}.

One of the major triumphs of diffusion models lies in the crystal structure generation, where material structures can generally be represented as $\textbf{M} = (\textbf{A}, \textbf{X}, \textbf{L})$, comprising atom types ($\textbf{A}$), atomic coordinates ($\textbf{X}$), and lattice matrices ($\textbf{L}$). Diffusion models have shown exceptional promise in jointly optimizing these components, with GNN frequently employed to capture the positional relationships between atoms\cite{xie2018crystal}. By incorporating symmetry constraints and periodic boundary conditions, diffusion models are particularly well-suited for generating stable and physically realizable crystal structures.
One pioneering framework is the crystal diffusion variational autoencoder (CDVAE), which combines VAE with a diffusion model to generate crystal structures\cite{xie2021crystal} (Fig.\,\ref{fig6}(b)). The workflow of CDVAE starts from generating the lattice unit cell, initial chemical composition and number of atoms using VAE. Then the diffusion model jointly updates atomic types and coordinates to finalize a crystal structure. By integrating diffusion processes with latent space representations, CDVAE facilitates diverse and chemically plausible material generation. Applications of CDVAE to 2D materials have demonstrated its capacity to generate novel stable structures with high chemical and structural diversity\cite{lyngby2022data}. Con-CDVAE realized the conditional generation of materials structures by carefully encoding the target properties like formation energy and bandgap, guiding the latent-based model to generate feasible structures with the target constraints\cite{ye2024cdvae}. Furthermore, conditional generation within latent spaces has also been employed for porous materials, where interpolation of latent vectors has uncovered novel zeolite structures optimized for properties like void fractions and adsorption capacity\cite{park2024inverse}. 

Most recently, diffusion models that operate directly on noisy material structures have gained attention for their simplicity and effectiveness in \textit{ab initio} structure generation, as they do not require sampling data from latent space prior to diffusion. 
Instead of fixing $\textbf{L}$ as in CDVAE, these models are initialized with randomized $(\textbf{A}, \textbf{X}, \textbf{L})$ and iteratively refine the structure to optimize $(\textbf{A}, \textbf{X}, \textbf{L})$ jointly. This makes them the most flexible framework for structure generation and materials design.
One example is DiffCSP, which utilizes periodic equivariant denoising processes to generate lattice and atomic coordinates jointly, ensuring symmetry constraints compliance and achieving superior performance in crystal structure generation\cite{jiao2023crystal}. 
Another architecture is MatterGen, which employs a flexible equivariant diffusion-based framework to generate crystalline structures, and doubles the stability and novelty rates compared to prior generative models. 
Furthermore, MatterGen can be fine-tuned with a small amount of labeled data to achieve conditional generation on various constraints, including chemical elements, symmetry and material properties like magnetic density and bulk modulus\cite{zeni2025generative} (Fig.\,\ref{fig6}(c)). Mattergen is also demonstrated to tackle materials design problems with multiple constraints, by searching for materials with high magnetic density and low supply-chain-risk.

Given the importance of generating particular geometrical patterns in quantum materials, structural constraint integration in generative model (SCIGEN) has been proposed, which imposes hard constraints on the crystal structure motif by combining constrained and unconstrained components during the diffusion process\cite{okabe2024structural}. 
As is shown in Fig.\,\ref{fig6}(d), the initialized structure of SCIGEN is subjected to a diffusion process by adding noise over $T$-steps denoted as $\boldsymbol{M}_{t}^c$, where $t \in [1..T]$, providing the pre-defined pathway of denoising process for the constrained components. 
The unconstrained structure is initiated as a completely noisy structure $\boldsymbol{M}_{T}^u$. Both $\boldsymbol{M}_{t}^c$ and $\boldsymbol{M}_{t}^u$ are integrated to form $\boldsymbol{M}_{t}$, which is then denoised to retrieve $\boldsymbol{M}_{t-1}^u$, the unconstrained structure of the previous time step. 
SCIGEN repeats this process through all steps until it optimizes the final material structure $\boldsymbol{M}_{0}$ with desired structure motif.
Utilizing this framework, SCIGEN generates millions of new compounds with various structural motifs such as honeycomb and kagome compounds (see Fig.\,\ref{fig6}(e) for a few examples).

Despite being state-of-the-art models for materials inverse design, diffusion models still have large room of improvement. A key limitation is the optimization of discrete components, such as the number of atoms per unit cell, which must remain consistent throughout the denoising process. Moreover, while the probablistic denoising process offers flexibility, it does not guarantee the stability or feasibility of interpolated structures, necessitating careful validation. Additionally, diffusion models demand significant computational resources due to their iterative nature, and their performance heavily depends on the quality and diversity of training data as well as the implementation details of the neural network architectures. Nevertheless, diffusion models have already demonstrated their transformative potential in inverse materials design.

We also briefly discuss flow matching\cite{lipman2022flow} and consistency models\cite{song2023consistency}, which are emerging generative models beyond diffusion models. Flow matching models directly learn continuous transformations between data distributions, offering efficient sampling and high-quality generation. Most recently, Luo \textit{et al.} proposed CrystalFlow, which achieves high performance across standard generation benchmarks and performs conditional generation for materials inverse design\cite{luo2024crystalflow}. Consistency model is the latest generative model architecture in computer vision, which achieves similar quality of data generation with much less sampling steps compared to diffusion models. This approach also has potential for accelerated materials generation and also for inverse design.

\subsection{Large language models (LLMs)}
LLMs arise from the concept of autoregressive models, which predict the next element in a sequence based on preceding elements. Powered by the transformer architecture with attention mechanism\cite{vaswani2017attention} and the GPT-2\cite{radford2019language}, LLMs with billions of parameters are advanced autoregressive transformers trained on vast textual datasets to generate coherent and contextually relevant text. Though originally designed for text autoregression, LLMs have been extensively applied in materials design. 
Specifically, researchers represent the structure of materials in a sequential text form and perform their tasks with LLMs accordingly. For example, CrystalFormer is an autoregressive transformer that generates crystals by leveraging space group symmetries, achieving efficient data usage and property-guided materials design\cite{cao2024space}; LLaMA-2 has been fine-tuned to generate stable inorganic materials, achieving higher metastable material generation rates and handling text-conditional prompts and infilling\cite{gruver2024fine}. Most recently, LLMs have also proved their immense power as agents optimizing material performance in an interactive style. To name a few, LLMatDesign, an LLM-driven framework for interactive materials discovery, enables iterative design and evaluation to achieve target properties like band gaps and stability\cite{jia2024llmatdesign}; dZiner uses a chemistry-informed AI agent powered by LLMs  toward inverse design. The agent leverages domain-specific insights to propose new compounds with desired properties, iteratively evaluating them using relevant surrogate models\cite{ansari2024dziner}. 
LLM has also been integrated with experiments, where it suggests possible synthesis recipe for horizontally aligned carbon nanotube (HACNT). Specifically, the AI-recommended TiPt catalyst outperforms traditional Fe catalysts, achieving higher-accuracy density control in HACNT arrays synthesis\cite{li2024transforming}.
With the rapid development of state-of-the-art LLMs towards artificial general intelligence (AGI)\cite{jaech2024openai}, LLMs are transforming the automated materials inverse design. However, rather than directly generating any material with desired properties or predicting any material properties, current LLMs seem to show more power as AI-agents that give suggestions than directly used for materials inverse design. 

\section{Future Perspectives}\label{future}
\begin{figure}[!htbp]
    \centering
    \includegraphics[width=0.9\linewidth]{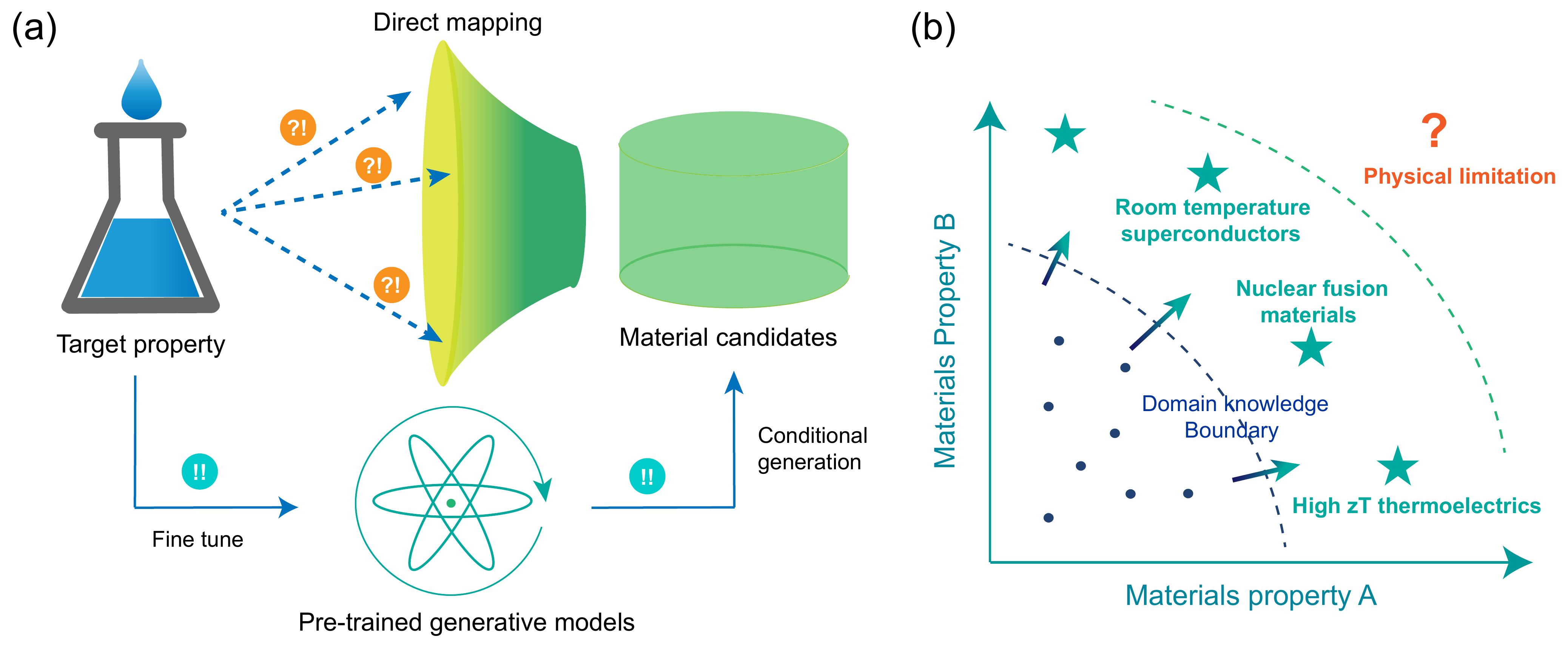}
    \caption{\textbf{Future perspectives of materials inverse design.} \textbf{(a)} Pre-trained generative models could be fine-tuned by target properties to perform conditional generation for inverse design, solving the ill-posed property-to-structure. \textbf{(b)} The paradigm of inverse design pushes the material space beyond our domain knowledge boundary, creating opportunities for discovering newzhoo material systems and exploring the limit of physical science.}
    \label{fig7}
\end{figure}

There is an ongoing paradigm shift in materials design, from forward screening within a fixed dataset to probabilistic inverse design, driven by the progress of deep generative models.
As shown in Fig.\,\ref{fig7}(a), conditional probabilistic generation is particularly important in inverse design because the problem lacks a one-to-one mapping between properties and structures and is ill-posed.
Therefore, the future of inverse design could lie in generative models, which excel in conditional generation and enable the design of materials with targeted, domain-specific properties.
These models offer ultrahigh flexibility and efficiency for navigating the complex design spaces of materials.
With a well-pretrained foundational generative model, one can in principle fine-tune the model conditioned on targeted properties with much sparser data\cite{zeni2025generative}, and perform inverse design at an unprecedented speed. 
Generative models shows great promise for multi-objective optimization tasks, such as the simultaneous improvement of thermoelectric efficiency\cite{yan2022high} or multiferroic coupling\cite{fiebig2016evolution}, which have been extremely challenging through traditional approaches.

Despite the potential and advanced techniques of generative models, inverse design still faces considerable challenges. 
One of the most pressing issues lies in ensuring the thermodynamic stability of predicted materials, which can be the prerequisite for synthesizability in labs. 
Current state-of-the-art generative models like MatterGen\cite{zeni2025generative} and DiffCSP\cite{jiao2023crystal} can generate millions of novel materials, yet verifying their stability with experimental feasibility remains challenging. 
Furthermore, the practical utility of these materials is another major challenge. While it has become feasible to generate millions of candidate materials, the ultimate goal is to identify a few ``holy-grail'' materials that have transformative properties -- possibly surpassing any known material -- for real-world applications. 
This is linked to another challenge, that to generate out-of-distribution (OOD) data. OOD is essential for discovering entirely new families of materials. 
Current generative models primarily focus on \textit{interpolation} within known datasets, limiting their ability to uncover groundbreaking materials such as room-temperature superconductors.  
Models capable of reliable extrapolation, at least to a certain level, are crucial for pushing the boundaries of materials science beyond the domain knowledge of human, as shown in Fig.\,\ref{fig7}(b).

Thousands of years have passed since humans began experimenting with bronze and iron. In the centuries-long pursuit of materials discovery, we have identified millions of materials, yet we are only at the dawn of the revolution.
Although computational materials discovery has greatly accelerated the search process in a larger design space, only a small fraction generated materials have been experimentally validated.
Looking ahead, current AI models should address the challenges regarding stability and domain knowledge extrapolation in order to fully realize the potential of conditional generation in inverse design. 
The integration of generative models into comprehensive workflows, combining efficient material discovery with robust experimental validation and AI-agents, could be promising for future materials design and close the design loop.   
By addressing stability, utility, and exploration challenges, inverse design powered by deep generative models can transform from a promising methodology to a reliable tool for scientific discovery and technological innovation.

\section*{Acknowledgements}
The authors thank Dr. Tian Xie for helpful discussions. MC and CF acknowledge support from the U.S. Department of Energy (DOE), Office of Science (SC), Basic Energy Sciences (BES), Award No. DE-SC0021940. RO and AC acknowledge support from National Science Foundation
(NSF) Designing Materials to Revolutionize and Engineer our Future (DMREF) Program with Award No. DMR-2118448. AB acknowledges the support from NSF ITE-2345084. ML acknowledges the Class of 1947 Career Development Chair, and the support from R. Wachnik.

\section*{Data availability}
This review does not involve the generation of new data.

\section*{Conflict of interest}
The authors declare no conflict of interest.

\bibliography{refs.bib} % Tell bibtex which .bib file to use

\end{document}